\documentclass[pra,aps,twocolumn,superscriptaddress,showpacs]{revtex4}
\usepackage{amsmath,bm,amssymb}
\usepackage{graphicx}
\newcommand{\bk}{b_{\bm{k}}}

\newcommand{\bkd}{b_{\bm{k}}^{\dag}}

\newcommand{\bmk}{b_{-\bm{k}}}

\newcommand{\ckl}{c_{\bm{k},\lambda}}

\newcommand{\ckld}{c_{\bm{k},\lambda}^{\dag}}

\newcommand{\sumk}{\sum_{\bm{k}}}
\newcommand{\sumkl}{\sum_{\bm{k},\lambda}}

\newcommand{\wk}{w_{\bm{k}}}

\newcommand{\omk}{\omega_{\bm{k}}}

\newcommand{\fk}{f_{\bm{k}}}

\newcommand{\gkl}{g_{\bm{k}\lambda}}

\newcommand{\kk}{\bm{k}}

\newcommand{\pe}{|1\rangle\!\langle 1|}

\newcommand{\be}{\langle 1|}
\newcommand{\bg}{\langle 0|}
\newcommand{\ke}{|1\rangle}
\newcommand{\kg}{|0\rangle}
\newcommand{\rl}{\rangle\!\langle}
\newcommand{\unit}{\mathbb{I}}
\DeclareMathOperator{\tr}{Tr}
\DeclareMathOperator{\re}{Re}

\begin{document}

\author{Pawe{\l} Machnikowski}
\email{pawel.machnikowski@pwr.wroc.pl}
\affiliation{Institute of Physics, Wroc{\l}aw University of
Technology, 50-370 Wroc{\l}aw, Poland}
\author{Katarzyna Roszak}
\affiliation{Institute of Physics, Wroc{\l}aw University of
Technology, 50-370 Wroc{\l}aw, Poland}
\affiliation{Department of Condensed Matter Physics,
Faculty of Mathematics and Physics, Charles University,
12116 Prague, Czech Republic}
\author{Anna Sitek}
\affiliation{Institute of Physics, Wroc{\l}aw University of
Technology, 50-370 Wroc{\l}aw, Poland}
\affiliation{Institut f{\"u}r Theoretische Physik, 
Nichtlineare Optik und Quantenelektronik, Technische Universit{\"a}t
Berlin, 10623 Berlin, Germany}

\title{Collective luminescence and phonon-induced processes in double quantum  
dots}

\begin{abstract}
We study the evolution of a quantum state of a double quantum dot
system interacting with the electromagnetic environment and with the
lattice modes, in the presence of a coupling between the two dots. We
propose a unified approach to the simulation of the system evolution
under joint impact of the two reservoirs. 
We discuss the sub- and superradiant radiative decay of the system, the
phonon-induced decay of entanglement between the dots, and the
transfer of excitation between them. 
\end{abstract}


\maketitle

\section{Introduction}

Semiconductor structures composed of two closely spaced quantum dots (QDs)
have attracted enormous attention in recent years. 
Pairs of vertically stacked 
QDs with spatial separation down to single nanometers can be obtained in
a two-layer self-assembled growth process, where the strain
distribution favors QD nucleation in the second layer on top of the
QDs formed in the first layer \cite{xie95,solomon96}.
The development of the manufacturing technologies
\cite{gerardot03,fafard99} has made it possible to achieve structures
built of nearly identical dots, with the splitting of ground state
transition energies down to single meV (very likely
interaction-limited) \cite{gerardot05,ortner05}. The state space of
such a double quantum dot (DQD) is obviously 
richer than that of a single QD \cite{szafran05,szafran08} and allows,
e.g., for entanglement 
between the dots. Also the recombination and relaxation processes in
DQDs show many features which cannot appear in individual QDs.
The quantum coherence of
carrier states in DQDs are affected by the interference and
collective  effects that appear in the interaction of such systems
with their radiative environment (electromagnetic vacuum) and with
the surrounding crystal lattice (phonons). 

Optical properties of DQDs may
be strongly modified due to collective interaction of sufficiently
closely spaced QDs with the electromagnetic (EM) field. 
These collective effects have been extensively
studied for atomic systems
\cite{gross82} where they 
manifest themselves by superradiant emission \cite{skribanovitz73}.
A signature of superradiant behavior was also observed in ensembles of
QDs \cite{scheibner07}.
On the other hand, the collective interaction leads
to the appearance of subradiant states which are decoupled from the
environment and, therefore, do not undergo decoherence. 
It has been proposed to use these states for
noiseless encoding of quantum information \cite{zanardi97,zanardi98b}. 

Another area where new features appear for DQD systems is the
decoherence due to carrier-phonon coupling. 
Dephasing of carrier
states in QDs due to carrier-phonon dynamics has been observed experimentally as a
decay of nonlinear optical response in a four-wave mixing experiment
with ultrashort pulses \cite{borri01,vagov04}.
A characteristic feature of the phonon-induced dephasing in QDs
is that it is only \textit{partial}, i.e., after a few picoseconds of
carrier-phonon dynamics, the degree of coherence (i.e., the values of
the off-diagonal elements of the density matrix) reaches a certain
finite level, depending on the system geometry and temperature
\cite{krummheuer02,jacak03b}. In DQDs and regular QD arrays, 
the degree of dephasing may be reduced
by encoding the logical qubit values into many-exciton states over a
QD array \cite{grodecka06}. Phonon-induced dephasing is also
detrimental to entanglement in DQDs and larger QD arrays. The impact
of the partial dephasing on the entanglement is very strong since the
latter is more prone to dephasing than local coherence and may be
completely destroyed even though the decoherence is only partial
\cite{diosi03,yu04,dodd04}. 
Moreover, because of the delicate
aspect of inter-subsystem coherences involved, the decay of
entanglement due to dephasing strongly depends on the nature of
environmental interaction (the same vs. different reservoirs for the
two subsystems) \cite{yu03}.

Various types of coupling between the dots change the properties of
the system even further. 
Theoretical
calculations show that for closely spaced dots, tunnel coupling (wave
function overlap) between the
dots should strongly affect their electronic structure
\cite{bryant93,schliwa01,szafran01}. Optical spectra
of such structures indeed show clear manifestations of electronic
coupling
\cite{ortner05,bayer01,ortner03,ortner05b,krenner05b}.

On the other hand, wave function overlap is not the only mechanism of
interaction between the QDs. In fact, for QD separations
of about 10~nm the energetically lowest states (in the
absence of external fields) correspond to spatially direct excitons
localized in individual QDs \cite{szafran01}. Such states are still bound by
Coulomb interaction. While the static (``direct'') dipole coupling preserves the
occupations of the individual QDs, the F\"orster interaction 
via interband dipole moments
\cite{lovett03b,ahn05} (first introduced in the context of molecular
systems \cite{forster48,dexter53}) makes it possible to 
transfer the exciton occupation between the dots. 
In contrast to the tunnel coupling, which is analogous to
chemical bonding between atoms and, therefore, turns the two dots into
one quantum system (a quantum dot molecule), 
the dipole couplings are rather like van der Waals
forces between separate entities. Therefore, QDs coupled by this kind
of interaction are much closer to the general paradigm of
well-defined, separate
qubits \cite{divincenzo00b} on which, however, collective quantum
operations (multi-qubit gates) can be performed, based on the
interactions between the systems.

In a closed system the (usually very weak) F{\"o}rster interaction has
considerable effects only very close to resonance
\cite{lovett03b,ahn05}. However, 
carrier-phonon coupling provides the necessary
dissipation channel which opens a possibility of excitation transfer
driven by the F{\"o}rster interaction even if the energy mismatch
between the dots is much larger than the interaction energy.
Phonon-assisted excitation transfer between the quantum states of a
molecule was in fact observed in many experiments 
\cite{gerardot05,heitz98,tackeuchi00,rodt03,ortner05c,nakaoka06}. 
In closely stacked
dots this excitation transfer process is mostly due to phonon-assisted
tunneling of carriers. However, for larger separations tunneling is
exponentially suppressed. Indeed, 
phonon-assisted transitions involving tunneling are very inefficient for a
10~nm separation even though a small energy splitting matches
the acoustic phonon energies \cite{nakaoka06}. In such cases, the
transfer is most likely due to the F\"orster coupling.

This paper presents a theoretical review of a few phenomena that appear
in double-dot systems. The existing results, which we originally
obtained using various theoretical approaches, are presented
within a unified formalism. Moreover, some examples of the interplay
between the effects due to the phonon and photon reservoirs are
discussed. In Sec.~\ref{sec:system} we define the model and our
approach to the simulation of its evolution. In
Sec.~\ref{sec:transfer}, we present a theory of excitation transfer between
coupled quantum dots
due to interband dipole coupling and 
carrier-phonon interaction \cite{rozbicki08a}. Next, in
Sec.~\ref{sec:radiative} we summarize our studies on collective
effects in spontaneous emission \cite{sitek07a}. We show that
superradiant-like behavior may 
appear in the optical response of sufficiently strongly coupled
pairs of quantum dots even if the two dots are not identical. 
In addition, we extend the existing result by
discussing the effect of phonon-induced transitions on the collective
optical properties. In Sec.~\ref{sec:entangle}, we
discuss the decay of entanglement due to phonon-induced
dephasing. Again, we extend the existing study of phonon-related
effects on picosecond time scales \cite{roszak06a} by
discussing the long-time decay due to spontaneous emission and the
mutual impact of phonon-related and radiation-related effects. The
final Sec.~\ref{sec:concl} contains concluding remarks and some
outlook for a possible further development of the theory.

\section{The system}
\label{sec:system}

The system under study is composed of two coupled QDs with transition energies
$\epsilon_{1}$ and $\epsilon_{2}$, 
interacting with their phonon and photon (radiative)
environments. We restrict the discussion to the ground states of
excitons in each dot and assume that the spin polarizations of the
excitons are fixed. As the exciton dissociation energy in absence of
external electric fields is rather large (several to a few tens of
meV), we consider only spatially direct exciton states, i.e., such
that the electron-hole pairs reside in one and the same dot.

Thus, the model includes four basis states $|mn\rangle$,
$m,n=0,1$, where $m$ and $n$ denote the number of excitons in the
first and second dot, respectively. 
We will use either the explicit tensor
product notation or a contracted one both for the states of the two dots
($|mn\rangle\equiv|m\rangle\otimes\ |n\rangle$) and for the operators
($|mn\rl m'n'|\equiv |m\rl m'|\otimes |n\rl n'|$).

We will describe the evolution in a ``rotating basis'' defined by the
unitary transformation 
\begin{equation*}
U=
e^{iE t(|01\rl 01|+|10\rl 10|+2|11\rl 11|)/\hbar},
\end{equation*}
where $E=(\epsilon_{1}+\epsilon_{2})/2$.
The Hamiltonian is then
\begin{equation*}
H=H_{\mathrm{DQD}}
+H_{\mathrm{ph}}+H_{\mathrm{rad}}
+H_{\mathrm{c-ph}}+H_{\mathrm{c-rad}}.
\end{equation*}

The first term describes exciton states in the DQD structure. 
\begin{eqnarray}\label{ham-DQD}
H_{\mathrm{DQD}} & = & \Delta(\pe\otimes\unit
-\unit\otimes\pe) \nonumber\\
&&+V(\ke\!\bg\otimes\kg\!\be+\mathrm{H.c}),
\end{eqnarray}
where $\Delta=(\epsilon_{1}-\epsilon_{2})/2$, $\mathbb{I}$ is the unit
operator, and $V$ is the coupling between the
dots, which can be assumed real. 
Electron and hole wave functions will be modelled by identical
anisotropic Gaussians 
with identical extensions $l$ in the $xy$ plane
and $l_{z}$ along $z$ for both particles,
\begin{displaymath}
\psi(\bm{r})\sim\exp\left( 
-\frac{1}{2}\frac{x^{2}+y^{2}}{l^{2}} - \frac{1}{2}\frac{z^{2}}{l_{z}^{2}}
 \right).
\end{displaymath}
The coupling $V$ may originate either from the Coulomb (F{\"o}rster)
interaction or from tunnel coupling.
In the former case it is related to the distance between the dots and
to band structure parameters \cite{sitek07a,rozbicki08a},
\begin{equation}\label{V}
V=\frac{e^{2}|a|^{2}}{4\pi\epsilon_{0}\epsilon_{\mathrm{r}}D^{3}}
f(D/l).
\end{equation}
Here $e$ is the electron charge, $\epsilon_{0}$ and 
$\epsilon_{\mathrm{r}}$ are the
vacuum permittivity and the dielectric constant of the semiconductor,
respectively, and 
\begin{displaymath}
a=\frac{\hbar P_{\mathrm{cv}}}{m_{0}E_{\mathrm{g}}}
\approx \frac{\hbar}{\sqrt{2m_{e}}E_{\mathrm{g}}},
\end{displaymath} 
where $P_{\mathrm{cv}}$ is the interband marix element of the momentum
operator, $m_{0}$ and $m_{e}$ are the free and effective electron
masses, and $E_{\mathrm{g}}$ is the band gap. The function $f(x)$
accounts for the correction to the point dipole approximation due to
the finite size of the dots (in fact, comparable to $D$) 
\cite{govorov05,rozbicki08a} and is given by \cite{machnikowski09a}
\begin{displaymath}
f(x) = \frac{x^{3}}{\sqrt{2\pi}}\int_{0}^{1}dt(1-t^{2})
\frac{u(t)-x^{2}t^{2}}{u^{5/2}(t)}
\exp\left[ -\frac{x^{2}t^{2}}{2u(t)} \right],
\end{displaymath}
where $u(t)=1-t^{2}+(l_{z}/l)^{2}t^{2}$.
This correction reduces the coupling for
$D\lesssim l$ and removes the $1/D^{3}$ singularity at $D\to 0$, while
it does not affect the coupling for $D\gg l$ since $f(x)\to 1$ as
$x\to\infty$. 
We will neglect the possible biexciton shift (coupling between static,
intraband dipole moments). 

The phonon modes are described by the free phonon Hamiltonian
\begin{displaymath}
H_{\mathrm{ph}}=\sumk\hbar\omk\bkd\bk,
\end{displaymath}
where $\bk,\bkd$ are bosonic
operators of the phonon modes and $\omk$ are the corresponding frequencies.
Interaction of carriers confined in the
DQD with phonons is modelled by the Hamiltonian
\begin{eqnarray}\label{ham-phon}
H_{\mathrm{c-ph}}&=&(\pe\otimes\unit)\sumk\fk^{(1)}(\bkd+\bmk) \\
&&+(\unit\otimes\pe)\sumk\fk^{(2)}(\bkd+\bmk),\nonumber 
\end{eqnarray}
where $\fk^{(1,2)}$
are system-reservoir coupling constants.
For Gaussian wave functions, the coupling constants for the
deformation potential coupling between 
confined charges and longitudinal phonon modes have the form 
$\fk^{(1,2)}=\fk e^{\pm ik_{z}D/2}$, where $D$ is
the distance between the subsystems and  
\begin{displaymath}
\fk=(\sigma_{\mathrm{e}}-\sigma_{\mathrm{h}})\sqrt{\frac{k}{2\varrho vc_{\mathrm{l}}}}
\exp\left[
-\frac{l_{z}^{2}k_{z}^{2}+l^{2}k_{\bot}^{2}}{4}\right].
\end{displaymath}
Here $v$ is the normalization volume,  
$k_{\bot/z}$ are momentum
components in the $xy$ plane and along the $z$ axis,
$\sigma_{\mathrm{e/h}}$ are deformation potential constants for
electrons/holes, $c_{\mathrm{l}}$ is the speed of longitudinal sound,
and $\varrho$ is the crystal density.

The third component in our modeling is the radiative reservoir (modes
of the electromagnetic field), described by the Hamiltonian
\begin{displaymath}
H_{\mathrm{rad}}=\sumkl\hbar\wk\ckld\ckl,
\end{displaymath}
where $\ckl,\ckld$ are photon creation and annihilation operators
and $\wk$ are the corresponding frequencies ($\lambda$ denotes
polarizations). 
The QDs
are separated by a distance much smaller than the relevant photon
wavelength $\lambda=2\pi\hbar c/E$, where
$E=(\epsilon_{1}-\epsilon_{2})/2$, so that the spatial dependence of
the EM field may be neglected (the Dicke limit). 
The Hamiltonian describing the
interaction of carriers with the 
EM modes in the dipole and rotating wave approximations is 
\begin{equation}\label{hI}
H_{\mathrm{c-rad}}  = 
\Sigma_{-}\sumkl e^{-iEt/\hbar}\gkl \ckld +\mathrm{H.c.},
\end{equation}
with 
\begin{displaymath}
\Sigma_{-}
=\kg\be\otimes\unit +\unit\otimes\kg\!\be
\end{displaymath}
and
\begin{displaymath}
\gkl=i\bm{d}\cdot\hat{e}_{\lambda}(\kk)
\sqrt{\frac{\hbar\wk}{2\varepsilon_{0}\varepsilon_{\mathrm{r}}v}},
\end{displaymath}
where 
$\bm{d}$ is the interband dipole moment ($d=ea$) and
$\hat{e}_{\lambda}(\kk)$ is the unit polarization vector
of the photon mode with the wave vector $\kk$ and polarization
$\lambda$.  
For wide-gap semiconductors with $E\sim 1$ eV, zero-temperature
approximation may be used for the radiation reservoir at any
reasonable temperature. 

In certain limiting cases, analytical formulas for the evolution of
the DQD system may be found. For uncoupled dots ($V=0$) interacting
only with lattice modes (phonons), an exact solution is available
\cite{roszak06a}. If only the radiative decay is included, a solution
in the Markov limit can be obtained \cite{sitek07a}. Here, we will use
a description which allows one to deal with the simultaneous action
of both these environments. We describe the evolution of the reduced
density matrix of the DQD system in the interaction
picture with respect to $H_{\mathrm{DQD}}$ by the equation
\begin{equation*}
\dot{\rho}=\mathcal{L}_{\mathrm{rad}}[\rho]
+\mathcal{L}_{\mathrm{ph}}[\rho].
\end{equation*}
Here the first term describes the effect of the radiative decoherence
in the Markovian limit in terms of the Lindblad dissipator
\begin{equation*}
\mathcal{L}_{\mathrm{rad}}[\rho]=\Gamma_{\mathrm{rad}} 
\left[ \Sigma_{-}(t)\rho\Sigma_{+}(t)
-\frac{1}{2}\{\Sigma_{+}(t)\Sigma_{-}(t),\rho \}_{+}\right],
\end{equation*}
where $\Sigma_{-}(t)=\Sigma_{+}^{\dag}(t)
=e^{iH_{\mathrm{DQD}}t/\hbar}\Sigma_{-}e^{-iH_{\mathrm{DQD}}t/\hbar}$
and 
\begin{displaymath}
\Gamma_{\mathrm{rad}}=\frac{E^{3}|\bm{d}|^{2}\sqrt{\epsilon_{\mathrm{r}}}}{
3\pi\epsilon_{0}c^{3}\hbar^{4}}
\end{displaymath}
is the spontaneous decay rate for a single dot. 
The second term accounts for the interaction with the non-Markovian
phonon reservoir. We use the time-convolutionless equation
\begin{equation}\label{tcl}
\mathcal{L}_{\mathrm{ph}}[\rho]=
-\int_{0}^{t}d\tau\tr_{\mathrm{ph}}\left[ 
H_{\mathrm{c-ph}}(t),\left[ H_{\mathrm{c-ph}}(\tau),\rho(t)\otimes\rho_{\mathrm{ph}} 
\right]  \right],
\end{equation}
where $H_{\mathrm{c-ph}}(t)
=e^{iH_{\mathrm{DQD}}t/\hbar}H_{\mathrm{c-ph}}e^{-iH_{\mathrm{DQD}}t/\hbar}$,
$\rho_{\mathrm{ph}}$ is the phonon density matrix at the thermal
equilibrium, and $\tr_{\mathrm{ph}}$ denotes partial trace with
respect to phonon degrees of freedom.

The above equation of motion for the reduced density matrix strictly
reproduces the results in the limiting cases mentioned
above. Moreover, for the case of a DQD coupled to phonons with
non-vanishing coupling $V$, it yields results reasonably close to
those obtained by a much more complex correlation expansion technique
\cite{rozbicki08a}. 

In numerical simulations, we take the parameters corresponding to a self-assembled
InAs/GaAs system: $\sigma_{\mathrm{e}}-\sigma_{\mathrm{h}}=9$~eV,
$\rho=5350$~kg/m$^{3}$, $c_{\mathrm{l}}=5150$~m/s, 
the wave function parameters
$l=4.5$~nm, $l_{z}=1$~nm,  and the radiative recombination time (for a
single dot) $1/\Gamma_{\mathrm{rad}}=400$~ps.

\section{Phonon-assisted excitation transfer}
\label{sec:transfer}

In this section we discuss the evolution of a DQD system coupled only
to its phonon reservoir. 
The system is then described by the Hamiltonian 
$H=H_{\mathrm{DQD}}+H_{\mathrm{ph}}+H_{\mathrm{c-ph}}$ 
[Eqs.~(\ref{ham-DQD}) and (\ref{ham-phon})]. 
We will assume that there is one exciton in the system,
initially localized in one of the dots. We will see that the interplay
of the coupling and phonon-assisted dissipation leads to irreversible
excitation transfer between the dots, which is due to a weak Coulomb
(F{\"o}rster) coupling between them.

\begin{figure}[tb]
\includegraphics[width=80 mm]{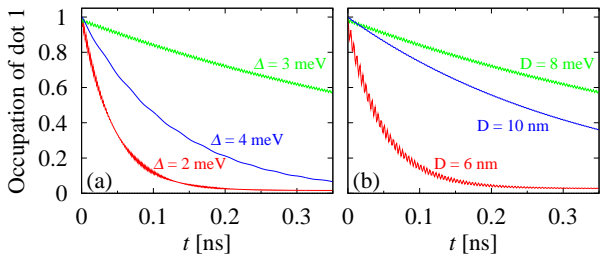}
\caption{The occupation of the higher energy QD as a function of time
at $T=4$~K: (a) $D=8$~nm and $\Delta$ as shown; (b) $\Delta=3$~meV
and $D$ as shown.}
\label{fig:kinet}
\end{figure}

In general, if the initial state corresponds to the exciton located in one
of the dots (state $|01\rangle$ or $|10\rangle$) the evolution is a
combination of conservative (unitary) oscillations due to the
perturbation induced by the coupling $V$ and a dissipative,
irreversible transition towards the lower eigenstate of $H_{\mathrm{DQD}}$
\cite{rozbicki07,rozbicki08a} induced by the coupling to the phonon
continuum. However, in the case of a weak coupling the oscillations
are very small and the occupation of the higher-energy dot follows a
nearly exponential decay, as shown in Fig.~\ref{fig:kinet}. 
In these computations, the coupling energy $V$ is calculated according
to Eq.~(\ref{V}).
An interesting feature visible in Fig.~\ref{fig:kinet} is that the
rate of the excitation transfer is non-monotonic both in the energy
mismatch and in the separation between the dots. 

The nearly exponential decay curve suggests that the process can be
described in the Markovian approximation. This is possible, since the
typical time scale of the decay process is long compared to the
transition frequency $V/\hbar$ between the two single-exciton states. We have shown
\cite{rozbicki08a} that the formal long-time limit of Eq.~(\ref{tcl})
in the rotating wave approximation
leads to optical Bloch equations describing the system dynamics in the
interaction picture and in the rotating frame related to the
eigenstates of $H_{\mathrm{DQD}}$. 

In general, upon transforming back to the original basis $|01\rangle,|10\rangle$
and to the Schr{\"o}dinger picture one obtains a complex
evolution. However, in the case of
$|V|\ll |\Delta|$, which is of particular practical importance, 
the eigenstates of $H_{\mathrm{DQD}}$ are very
close to $|10\rangle$ and $|01\rangle$.
Moreover, in this limit the energy difference between the eigenstates,
$\hbar\Omega=2\sqrt{V^{2}+\Delta^{2}}$
is nearly equal to $2\Delta$. In this limit, one obtains an
exponential excitation transfer with the rate 
\begin{displaymath}
\Gamma=4\pi\left(\frac{\tilde{V}}{\Delta}\right)^{2}
\left[R(\Omega)+R(-\Omega)\right],
\end{displaymath}
with the spectral density
\begin{displaymath}
R(\omega)=\frac{2}{\hbar^{2}}\left| n_{\mathrm{B}}(\omega)+1\right|
\sumk \sin^{2}\frac{k_{z}D}{2}\left| \fk\right|
\delta(|\omega|-\wk).
\end{displaymath}
Thus, the Markovian
equations are particularly 
useful in the limit of weak coupling, where the Markovian dephasing
rate may be identified with the rate of irreversible excitation
transfer between the dots. 

\begin{figure}[tb]
\includegraphics[width=83 mm]{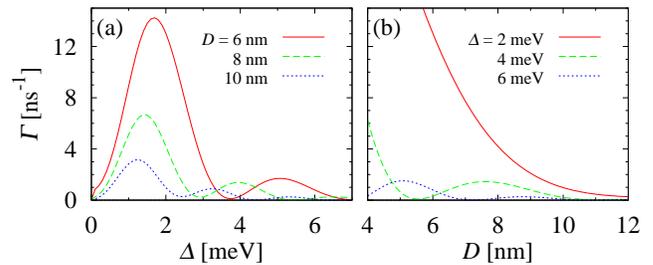}
\caption{The rate of phonon-assisted excitation transfer 
as a function of the energy mismatch for a few values of
the QD separation $D$ (a) and as a function of $D$ for a few values of
$\Delta$ (b) at $T=4$~K.}
\label{fig:rate}
\end{figure}

The rate for the phonon-assisted process is governed, on the one hand, by
the amplitude of the 
F{\"o}rster coupling which decreases roughly as $1/D^{3}$. On the
other hand, it is strongly influenced by the structure of
$R(\Omega)$. The latter oscillates as a function of both $\Omega$ and
$D$ \cite{rozbicki08a} due to the interplay between the wavelength of
the emitted phonon 
and the QD separation in the molecule (phonons are
preferentially emitted along the strongest confinement limit, i.e.,
along the DQD axis).
In particular, $R(\Omega)$ has a pronounced minimum whenever 
$\Omega$ is a multiple of $2\pi\hbar c_{\mathrm{l}}/D$ which explains the 
oscillating dependence on both $\Delta$ and $D$ (see Fig.~\ref{fig:rate}). 

The transfer rate for $\Delta=4$~meV,
$D=4.5$~nm is $\Gamma=2.3$~ns$^{-1}$, which is about twice lower than the
value of $\Gamma=5.25$~ns$^{-1}$ deduced from fitting to the photon
correlation data \cite{gerardot05}. It should be noted, however, that
our modelling is based on certain choices of parameters that cannot
be uniquely determined. First of all, the magnitude of the
F\"orster coupling for a given inter-dot distance 
can only be roughly estimated, since the value of the parameter $a$ in
an inhomogeneous, strained structure is not exactly known. Moreover, the
carrier-phonon coupling constants in the relevant energy (or wave
vector) range are strongly geometry dependent. For instance, by changing
the localization widths to $l_{\mathrm{e}}=l_{\mathrm{h}}=4.0$~nm and
$l_{z}=0.8$~nm one gets a considerably increased value of
$\Gamma=3.65$~ns$^{-1}$. 

\section{Collective radiative properties of double quantum dots}
\label{sec:radiative}

In this section,  we study the effect of the collective coupling
between the two interacting QDs and their electromagnetic environment
on the radiative relaxation of carriers. 
We begin with a model of a DQD interacting only with
the EM field. Then, we discuss the effect of the additional
phonon-induced relaxation.

In the case of purely electromagnetic environment, the Hamiltonian of
the system is  $H=H_{\mathrm{DQD}}+H_{\mathrm{rad}}+H_{\mathrm{c-rad}}$.
This Hamiltonian conserves the number of excitations (excitons plus
photons).  In this case, the solution obtained using the equation
(\ref{tcl}) is equivalent to that derived using the Wigner--Weisskopf
approach \cite{sitek07a}.

\begin{figure}[tb]
\begin{center}
\unitlength 1mm
\begin{picture}(85,30)(0,11)
\put(0,5){\resizebox{80mm}{!}{\includegraphics{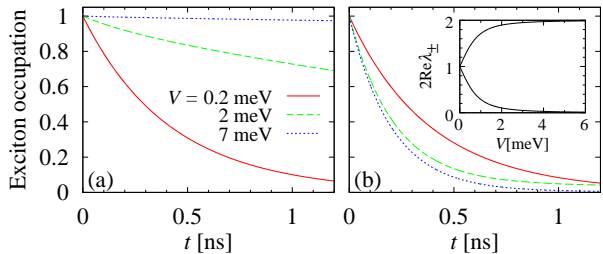}}}
\end{picture}
\end{center}
\caption{\label{fig:occup1meV}The exciton occupation for
sub- (a) and superradiant (b) states for $\Delta=1$ meV. The inset in
(b) shows the values of the occupation decay rates. Line definitions
in (a) refer to both plots.}
\end{figure}

In Fig.~\ref{fig:occup1meV} we show the evolution of  the average number of excitons in the
DQD  for two initial states 
\begin{displaymath}
|\psi_{\pm}(0)\rangle=(|01\rangle\pm |10\rangle)/\sqrt{2},
\end{displaymath}
for a DQD with
a realistic value of the energy mismatch $\Delta=1$ meV. 
For $V\ll\Delta$, both states show simple exponential decay with the rate
$\Gamma_{\mathrm{rad}}$. This is understandable, since two different dots emit
radiation into different frequency sectors of the reservoir and no
collective effects should be expected. The situation changes 
in the opposite limit, $V\gg\Delta$. Now, one of the states
becomes stable (subradiant), while the other (superradiant) state
decays exponentially with 
a twice larger rate. It should be noted that the energy mismatch is
much larger than the emission line width, so that the subradiance and
superradiance effect is due to the coupling between the dots.
In the intermediate range of parameters, the
decay is not exponential. It can be shown \cite{sitek07a} that for the
superradiant initial state, the number of  
excitons evolves as 
\begin{eqnarray*}
n(t)=|\bm{c}(t)|^{2} & = & \sin^{2}(\varphi+\pi/4)
e^{2\re\lambda_{-}t}\\
&&+\cos^{2}(\varphi+\pi/4)e^{2\re\lambda_{+}t},
\end{eqnarray*} 
where 
\begin{displaymath}
\sin\varphi= \frac{1}{\sqrt{2}}\left( 
1-\frac{\Delta}{\sqrt{\Delta^{2}+V^{2}}}\right)^{1/2}
\end{displaymath}
and
\begin{displaymath}
\lambda_{\pm}=-\frac{\Gamma_{\mathrm{rad}}}{2}\pm
\sqrt{-\Delta^{2}+(iV+\Gamma_{\mathrm{rad}}/2)^{2}}.
\end{displaymath}
The values of the two exponents $\lambda_{\pm}$ are shown in the inset
of Fig.~\ref{fig:occup1meV}(b).

\begin{figure}[tb]
\begin{center}
\unitlength 1mm
\begin{picture}(85,30)(0,11)
\put(0,5){\resizebox{80mm}{!}{\includegraphics{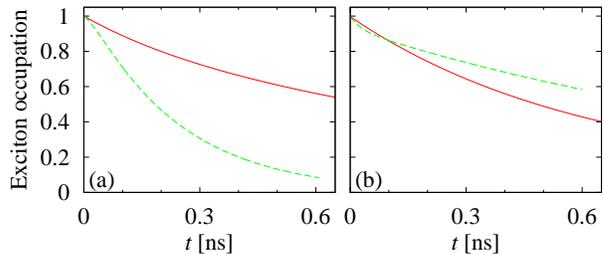}}}
\end{picture}
\end{center}
\caption{\label{fig:occup1meV-ph}The exciton occupation for
sub- and superradiant states in the presence of phonon-induced
relaxation for $\Delta=1$~meV, $d=8$~nm, $T=4$~K. (a) Subradiant state for
$V=-1$~meV; (b) superradiant state, $V=1$~meV. Red solid lines show
the evolution without phonons and the green dashed lines with phonons.}
\end{figure}

The subradiance and superradiance effects discussed above appear in spite of
the large energy mismatch because the single-exciton eigenstates of
the DQD Hamiltonian are superpositions of the basis states
$|01\rangle$ and $|10\rangle$ which are partly of sub- or superradiant
character (for the $-1$ and $+1$ relative phase, respectively). It is
therefore clear that the stability of these states is essential for
the effect. 

When carrier-phonon interaction is included, these states
are no longer stable due to excitation transfer processes discussed in
Sec.~\ref{sec:transfer}. Now, the exciton population decay will depend on
the interplay of radiative and phonon-related effects. This is
illustrated in Fig.~\ref{fig:occup1meV-ph}, where we plot the exciton
number as a function of time for the same two initial states as
previously, but now we include the coupling to phonons. Contrary to
the purely optical case, the kinetics now depends on the sign of the
coupling. For $V<0$, the lowest single-exciton state has a
superradiant character. Due to phonon-induced relaxation, the system
undergoes transition to this state and the subradiance effect is
partly destroyed [see Fig.~\ref{fig:occup1meV-ph}(a)]. An opposite
situation takes place for $V>0$ [Fig.~\ref{fig:occup1meV-ph}(b)]. Now,
the phonon-induced relaxation consists in a transition to the
subradiant state. As a result, the radiative recombination is slowed
down by the additional phonon-related decoherence.

Let  us note that the superradiant state is particularly
relevant for optical experiments since such a bright combination of
single exciton states is excited by ultrafast optical pulses from the
ground state. Thus, the curves plotted in
Figs.~\ref{fig:occup1meV}(b) and \ref{fig:occup1meV-ph}(b) directly
correspond to the decay of  population after an optical excitation.

\section{Entanglement decay}
\label{sec:entangle}

In this section we discuss the evolution of entanglement between the
states of two QDs under dephasing caused by carrier-phonon
and carrier-photon interactions. In our earlier work \cite{roszak06a}
we studied two dots in the absence of excitation transfer coupling,
undergoing only phonon-induced pure dephasing. Since that problem was
represented by an independent boson model the evolution of the open system
could be found in a closed analytical form. We showed that
phonon-induced pure dephasing, in spite of its only partial character,
can lead to complete decay of entanglement after a finite time. As we
pointed out, this happens for certain initial
maximally entangled states in which all the coherences are present,
that is, all four basis states are involved in the superposition. 
We studied also the dependence of the entanglement decay on
the distance between the dots. We showed that the degree of
dephasing-induced disentanglement increases with growing distance
between the dots and that non-zero distance is a necessary condition
for complete disentanglement. As expected, the degree of dephasing
and, therefore, disentanglement increases as the temperature grows.
The decay of entanglement due to coupling with the electromagnetic
field was also studied \cite{yu04}.

Here, we extend our earlier model \cite{roszak06a} by including the
radiative decay 
of the entangled excitons (coupling to the photon reservoir) and
excitation transfer interaction ($V\neq 0$) between the dots. Since
the extended model becomes more complicated and presents many new
features, we focus on these new elements and restrict the discussion
to one initial state and fixed distance $D=8$~nm between the dots. We
assume also that the biexciton shift is absent. The dots are assumed
different, with the energy mismatch of $\Delta=1$~meV.
In the presence of spontaneous emission and inter-dot coupling,
the evolution equation cannot be solved
analytically and we use the numerical approach described in
Sec.~\ref{sec:system}. We study the evolution of a maximally entangled
``singlet'' state 
\begin{displaymath}
|\psi(t=0)\rangle=\frac{|01\rangle-|10\rangle}{\sqrt{2}}.
\end{displaymath}
As a measure of entanglement, we use the Wootters concurrence
\cite{hill97,wootters98}. 

\begin{figure}[tb] 
\begin{center} 
\unitlength 1mm
\begin{picture}(85,30)(0,6)
\put(0,0){\resizebox{80mm}{!}{\includegraphics{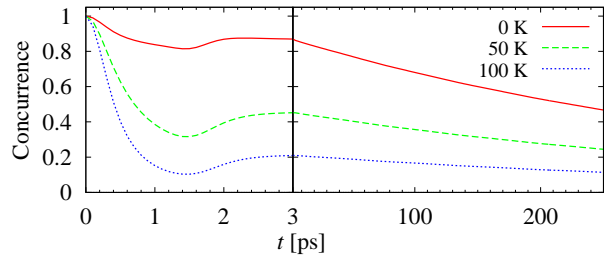}}}
\end{picture} 
\end{center} 
\caption{\label{fig:entang}
Evolution of entanglement for uncoupled dots ($V=0$) at various
temperatures.}
\end{figure}

The evolution of the entanglement between the dots is shown in
Fig.~\ref{fig:entang}. At short (picosecond) time scales, phonon-induced
dephasing leads to a drop of concurrence, ending with a temperature-dependent
plateau level. In the absence of radiative recombination, the
entanglement would remain constant after this initial dephasing
stage. However, in the presence of carrier-photon coupling the exciton
lifetime becomes finite, which leads to an exponential population
decay and, in consequence, to decay of entanglement on the time scales
$\sim 1/\Gamma_{\mathrm{rad}}$.

\begin{figure}[tb] 
\begin{center} 
\unitlength 1mm
\begin{picture}(85,30)(0,6)
\put(0,0){\resizebox{80mm}{!}{\includegraphics{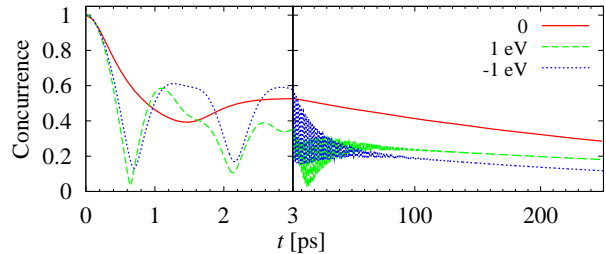}}}
\end{picture} 
\end{center} 
\caption{\label{fig:entang2}
Evolution of entanglement for coupled dots ($V=\pm 1$~meV, as shown)
compared to the uncoupled case ($V=0$) at $T=40$~K.}
\end{figure}

The situation becomes much more complicated (and more interesting) if
the dots are coupled by a transfer-type interaction (that is, $V\neq
0$). The evolution in this case is shown in Fig.~\ref{fig:entang2}. A
few effects can be seen. The most striking feature are the
oscillations of concurrence. Since the coupling is comparable to the
energy mismatch, the system performs rotations in the single-exciton
subspace, coming close to the separable states $|01\rangle$ or
$|10\rangle$ every half-period. These oscillations are damped on a
time scale of tens of picoseconds, as the excitation is dissipatively
transferred to the lower-energy eigenstate of $H_{\mathrm{DQD}}$ by a
process discussed above in Sec.~\ref{sec:transfer}. Depending on the
sign of the interaction, this eigenstate (which can still be
entangled) can have either subradiant or superradiant character (for
$V>0$ and $V<0$, respectively). This is visible as the difference in
the entanglement decay rates between the two cases for long times.

\section{Conclusions and outlook}
\label{sec:concl}

In this paper, we have reviewed and extended our recent results on the
interaction 
between systems composed of two quantum dots and their
environment. We have shown that collective interaction between the
dots and the surrounding radiation reservoir leads to sub- and
superradiance effects. In the presence of a coupling between the
dots, these collective effects become more stable
against the differences between the two transition energies. We have also
seen that carrier-phonon coupling leads to new dephasing effects
in such double-dot systems. In particular, phonon-induced decoherence
destroys entanglement between the two dots. This destruction of
entanglement is much stronger than local dephasing. We have
discussed the joint effect of phonon-induced dephasing and inter-dot
coupling. In this case, the most important feature of the system
dynamics is the irreversible excitation transfer between the dots.

A new area of interesting and important phenomena emerges if both the
spontaneous emission and phonon-induced dephasing are simultaneously
taken into account. Most of this field remains to be exploited. As an
example of physical effects that may appear due to joint impact of the
two reservoirs we have discussed the phonon-induced modification of
collective emission due to transitions between subradiant and
superradiant states and the interplay of the two dephasing channels in
the decay of entanglement between excitons confined in the two dots.

Obviously, the phenomena discussed here are only a fraction of all the
physical effects that may be present in a system of two non-identical
quantum dots interacting with two different reservoirs (photons and
phonons) in the presence of coupling between the dots. In particular,
the evolution 
of double quantum dot systems undergoing simultaneous dephasing via
both radiative and phonon-related channel has been studied only in a
very limited scope. 
Investigation
of these problems will certainly bring new knowledge not only on the
properties of quantum coherence in this specific semiconductor system,
but also on general properties of open quantum systems.

\begin{acknowledgments}
This work was supported in parts by the Polish MNiSW (Grant No. N N202
105236) and by the Czech Science Foundation (Grant No. 202/07/J051).  
\end{acknowledgments}


\begin{thebibliography}{10}

\bibitem{xie95}
Q. Xie, A. Madhukar, P. Chen, and N.~P. Kobayashi, Phys. Rev. Lett. {\bf 75},
  2542  (1995).

\bibitem{solomon96}
G.~S. Solomon, J.~A. Trezza, A.~F. Marshall, and J.~S. Harris, Jr., Phys. Rev.
  Lett. {\bf 76},  952  (1996).

\bibitem{gerardot03}
B. Gerardot, I. Shtrichman, D. Hebert, and P. Petroff, J. Cryst. Growth {\bf
  252},  44  (2003).

\bibitem{fafard99}
S. Fafard, Z.~R. Wasilewski, C.~N. Allen, D. Picard, M. Spanner, J.~P.
  McCaffrey, and P.~G. Piva, Phys. Rev. B {\bf 59},  15368  (1999).

\bibitem{gerardot05}
B.~D. Gerardot, S. Strauf, M.~J.~A. de~Dood, A.~M. Bychkov, A. Badolato, K.
  Hennessy, E.~L. Hu, D. Bouwmeester, and P.~M. Petroff, Phys. Rev. Lett. {\bf
  95},  137403  (2005).

\bibitem{ortner05}
G. Ortner, I. Yugova, G. {Baldassarri H{\"o}ger von H{\"o}gersthal}, A.
  Larionov, H. Kurtze, D.~R. Yakovlev, M. Bayer, S. Fafard, Z. Wasilewski, P.
  Hawrylak, Y.~B. Lyanda-Geller, T.~L. Reinecke, A. Babinski, M. Potemski,
  V.~B. Timofeev, and A. Forchel, Phys. Rev. B {\bf 71},  125335  (2005).

\bibitem{szafran05}
B. Szafran, T.~C. F.~M. Peeters, S. Bednarek, J. Adamowski, and B. Partoens,
  Phys. Rev. B {\bf 71},  205316  (2005).

\bibitem{szafran08}
B. Szafran, Acta Phys. Polon. A {\bf 114},  1013  (2008).

\bibitem{gross82}
M. Gross and S. Haroche, Phys. Rep. {\bf 93},  301  (1982).

\bibitem{skribanovitz73}
N. Skribanowitz, I.~P. Herman, J.~C. MacGilvray, and M.~S. Feld, Phys. Rev.
  Lett. {\bf 30},  309  (1973).

\bibitem{scheibner07}
M. Scheibner, T. Schmidt, L. Worschech, A. Forchel, G. Bacher, T. Passow, and
  D. Hommel, Nature Physics {\bf 3},  106  (2007).

\bibitem{zanardi97}
P. Zanardi and M. Rasetti, Phys. Rev. Lett. {\bf 79},  3306  (1997).

\bibitem{zanardi98b}
P. Zanardi and F. Rossi, Phys. Rev. Lett. {\bf 81},  4752  (1998).

\bibitem{borri01}
P. Borri, W. Langbein, S. Schneider, U. Woggon, R.~L. Sellin, D. Ouyang, and D.
  Bimberg, Phys. Rev. Lett. {\bf 87},  157401  (2001).

\bibitem{vagov04}
A. Vagov, V.~M. Axt, T. Kuhn, W. Langbein, P. Borri, and U. Woggon, Phys. Rev.
  B {\bf 70},  201305(R)  (2004).

\bibitem{krummheuer02}
B. Krummheuer, V.~M. Axt, and T. Kuhn, Phys. Rev. B {\bf 65},  195313  (2002).

\bibitem{jacak03b}
L. Jacak, P. Machnikowski, J. Krasnyj, and P. Zoller, Eur. Phys. J. D {\bf 22},
   319  (2003).

\bibitem{grodecka06}
A. Grodecka and P. Machnikowski, Phys. Rev. B {\bf 73},  125306  (2006).

\bibitem{diosi03}
L. Di\'osi,  in {\em Irreversible Quantum Dynamics (Lecture Notes in Physics
  vol. 622)}, edited by F. Benatti and R. Floreanini (Springer, Berlin, 2003),
  pp.\ 157--163, quant-ph/0301096.

\bibitem{yu04}
T. Yu and J.~H. Eberly, Phys. Rev. Lett. {\bf 93},  140404  (2004).

\bibitem{dodd04}
P.~J. Dodd and J.~J. Halliwell, Phys. Rev. A {\bf 69},  052105  (2004).

\bibitem{yu03}
T. Yu and J.~H. Eberly, Phys. Rev. B {\bf 68},  165322  (2003).

\bibitem{bryant93}
G.~W. Bryant, Phys. Rev. B {\bf 47},  1683  (1993).

\bibitem{schliwa01}
A. Schliwa, O. Stier, R. Heitz, M. Grundmann, and D. Bimberg, Phys. Stat. Sol.
  (b) {\bf 224},  405  (2001).

\bibitem{szafran01}
B. Szafran, S. Bednarek, and J. Adamowski, Phys. Rev. B {\bf 64},  125301
  (2001).

\bibitem{bayer01}
M. Bayer, P. Hawrylak, K. Hinzer, S. Fafard, M. Korkusinski, Z.~R. Wasilewski,
  O. Stern, and A. Forchel, Science {\bf 291},  451  (2001).

\bibitem{ortner03}
G. Ortner, M. Bayer, A. Larionov, V.~B. Timofeev, A. Forchel, Y.~B.
  {Lyanda-Geller}, T.~L. Reinecke, P. Hawrylak, S. Fafard, and Z. Wasilewski,
  Phys. Rev. Lett. {\bf 90},  086404  (2003).

\bibitem{ortner05b}
G. Ortner, M. Bayer, Y. Lyanda-Geller, T.~L. Reinecke, A. Kress, J.~P.
  Reithmaier, and A. Forchel, Phys. Rev. Lett. {\bf 94},  157401  (2005).

\bibitem{krenner05b}
H.~J. Krenner, M. Sabathil, E.~C. Clark, A. Kress, D. Schuh, M. Bichler, G.
  Abstreiter, and J.~J. Finley, Phys. Rev. Lett. {\bf 94},  057402  (2005).

\bibitem{lovett03b}
B.~W. Lovett, J.~H. Reina, A. Nazir, and G.~A.~D. Briggs, Phys. Rev. B {\bf
  68},  205319  (2003).

\bibitem{ahn05}
K.~J. Ahn, J. F{\"o}rstner, and A. Knorr, Phys. Rev. B {\bf 71},  153309
  (2005).

\bibitem{forster48}
T. F{\"o}rster, Ann. Phys. (Leipzig) {\bf 2},  55  (1948).

\bibitem{dexter53}
D.~L. Dexter, J. Chem. Phys. {\bf 21},  836  (1953).

\bibitem{divincenzo00b}
D.~P. {DiV}incenzo, Fortschr. Phys, {\bf 48},  771  (2000).

\bibitem{heitz98}
R. Heitz, I. Mukhametzhanov, P. Chen, and A. Madhukar, Phys. Rev. B {\bf 58},
  R10151  (1998).

\bibitem{tackeuchi00}
A. Tackeuchi, T. Kuroda, K. Mase, Y. Nakata, and N. Yokoyama, Phys. Rev. B {\bf
  62},  1568  (2000).

\bibitem{rodt03}
S. Rodt, V. Turck, R. Heitz, F. Guffarth, R. Engelhardt, U.~W. Pohl, M.
  Stra{\ss}burg, M. Dworzak, A. Hoffmann, and D. Bimberg, Phys. Rev. B {\bf
  67},  235327  (2003).

\bibitem{ortner05c}
G. Ortner, R. Oulton, H. Kurtze, M. Schwab, D.~R. Yakovlev, M. Bayer, S.
  Fafard, Z. Wasilewski, and P. Hawrylak, Phys. Rev. B {\bf 72},  165353
  (2005).

\bibitem{nakaoka06}
T. Nakaoka, E.~C. Clark, H.~J. Krenner, M. Sabathil, M. Bichler, Y. Arakawa, G.
  Abstreiter, and J.~J. Finley, Phys. Rev. B {\bf 74},  121305(R)  (2006).

\bibitem{rozbicki08a}
E. Rozbicki and P. Machnikowski, Phys. Rev. Lett. {\bf 100},  027401  (2008).

\bibitem{sitek07a}
A. Sitek and P. Machnikowski, Phys. Rev. B {\bf 75},  035328  (2007).

\bibitem{roszak06a}
K. Roszak and P. Machnikowski, Phys. Rev. A {\bf 73},  022313  (2006).

\bibitem{govorov05}
A.~O. Govorov, Phys. Rev. B {\bf 71},  155323  (2005).

\bibitem{machnikowski09a}
P. Machnikowski and E. Rozbcki, Phys. Stat. Sol. (b) {\bf 246},  320  (2009).

\bibitem{rozbicki07}
E. Rozbicki and P. Machnikowski, Acta Phys. Pol. A {\bf 112},  197  (2007).

\bibitem{wootters98}
W.~K. Wootters, Phys. Rev. Lett. {\bf 80},  2245  (1998).

\bibitem{hill97}
S. Hill and W.~K. Wootters, Phys. Rev. Lett. {\bf 78},  5022  (1997).

\end{thebibliography}

\end{document}